\documentclass[12pt]{iopart}
\usepackage{amsmath}
\usepackage{amssymb}
\usepackage{cite}
\usepackage{graphicx}
\usepackage{amsfonts}

\begin{document}

\title[Optical OAM communication through 1\,km horizontal path]{Simulating
thick atmospheric turbulence in the lab with application to orbital angular
momentum communication}

\author{Brandon~Rodenburg,$^{1}$ Mohammad~Mirhosseini,$^{1}$ Mehul~Malik,$^{1}$ Omar~S.~Maga\~na-Loaiza,$^1$ Michael~Yanakas,$^{1}$ Laura~Maher,$^{1}$ Nicholas~K.~Steinhoff,$^{2}$ Glenn~A.~Tyler,$^{2}$ and~Robert~W.~Boyd$^{1,3}$}
\address{
$^1$The Institute of Optics, University of Rochester, Rochester, NY 14627 USA
\\
$^2$The Optical Sciences Company, 1341 South Sunkist Street, Anaheim, California 92806
\\
$^3$Department of Physics, University of Ottawa, Ottawa ON K1N 6N5 Canada
}

\ead{Brandon.Rodenburg@gmail.com} 

\begin{abstract}
We describe a procedure by which a long ($\gtrsim 1\,\mathrm{km}$) optical path
through atmospheric turbulence can be experimentally simulated in a controlled
fashion and scaled down to distances easily accessible in a laboratory setting.
This procedure is then used to simulate a 1-km-long free-space communication
link in which information is encoded in orbital angular momentum (OAM) spatial
modes.  We also demonstrate that standard adaptive optics methods can be used
to mitigate many of the effects of thick atmospheric turbulence.
\end{abstract}


\submitto{\NJP}
\maketitle

\section{Introduction}
Understanding how light propagates is one of the most fundamental topics in the
field of optics. Although this problem is very well understood for
deterministic systems, the problem of understanding how light behaves in random
or fluctuating media is still a very active area of
research~\cite{Andrews2005,Popoff2010,Vellekoop2010a,Metzger2013}.  Even
propagation through the air, which at first glance might be thought of as being
equivalent to free-space propagation, will show stochastic behavior when
observed over sufficiently long distances due to small random fluctuations in
the refractive index along the path.

Early research studying optical propagation along random paths arose in the
context of imaging of astronomical objects through the turbulence in
atmosphere~\cite{Young1974}. The most important effect on image quality in such
systems is the random phase imprinted onto the beam by the turbulence.  This
phase aberration can be described by the quantity $r_0$, a coherence length
scale of the turbulence defined in the receiver aperture and known as Fried's
parameter~\cite{Fried1965,Fried1968}. If all other effects of turbulence can be
ignored (e.g. amplitude fluctuations), then the turbulence can be approximated
by a random phase screen in the aperture of the receiver.  This approximation,
known as the ``thin phase screen approximation,'' simplifies the problem and
allows the turbulence strength to be fully characterized by the dimensionless
parameter, $D/r_0$, where $D$ is the diameter of the aperture or beam. Thus the
effects of turbulence depend not only on the intrinsic fluctuations in the air,
but also on details of the system.  This thin screen approximation is often
appropriate in astronomical systems as turbulence effects on beam propagation
are greatest where the atmosphere is thickest, which is typically located
directly in front of the telescope.

Another area in which understanding the effects of turbulence on optical beams
is important is in optical communication in free-space~\cite{Wang2012}.  It is
well known that beams of light that contain a phase vortex of the form
$\psi(r,\phi)\propto\exp{(i\ell\phi)}$, where $\ell$ is an integer, carry
orbital angular momentum (OAM).  A single photon prepared in such a state will
have an OAM equal to $\ell\hbar$ in addition to any angular momentum carried by
the polarization~\cite{Allen1992}.  In recent years there has been a good deal
of excitement based on encoding information in a free-space channel onto such
OAM beams of light~\cite{Gibson2004, Wang2012, Boyd2011a}, as such modes are a
natural basis for such systems (i.e. they are the eigenmodes of a channel with
cylindrical symmetry)~\cite{Mirhosseini2013}.  Such spatial mode encoding
schemes allow for increasing of the bit rate as each pulse may contain more
than 2 possible symbols.  This scheme also provides an enhancement to the
security when used for quantum key distribution~\cite{Bourennane2002,Cerf2002}.

Degradation of the signal in a free-space channel due to atmospheric turbulence
is a primary limitation to the information-carrying capacity of such a channel.
Much work has been done to study how this affects communication systems that
utilize spatial mode
encoding~\cite{Paterson2005,Tyler2009,Rodenburg2012,Gbur2008,Malik2012,Roux2011,Boyd2011};
however much of this work focuses on turbulence that is fully described by a
random thin phase screen in the receiver aperture. In more realistic
situations, such as communication along a long horizontal path through which
the turbulence is continuously distributed, one will see amplitude
fluctuations, that is scintillation, in addition to the problems caused by pure
phase fluctuations.  Some of this degradation can still be compensated for by
phase-only adaptive optics (AO), correcting for low-spatial-frequency
aberrations for a horizontal path of a few kilometers or
more~\cite{Levine1998}.  For more stronger scintillation one will begin to see
intensity nulls that are associated with phase
vortices~\cite{Fried1992,Tyler2000}.  These phase vortices, or branch points,
are known to degrade the performance of AO systems~\cite{Fried1998}, and there
is a complete breakdown in the performance for horizontal paths greater than
approximately 5\,km due to this effect~\cite{Primmerman1995}.  For
communication systems that communicate using OAM, this phenomenon presents an
additional problem as phase vortices are precisely the means of the encoding,
and randomly generated vortices introduce errors into such a scheme.

The propagation of an optical beam through thick turbulence is not in general
analytically solvable and thus requires either simulation or testing in a real
world setup.  The cost and lack of control of testing in a real-world setup
makes finding suitable methods of simulating turbulence highly desirable for
understanding this problem. In this paper we describe a method of simulating a
thick turbulence channel and show how this can be implemented in a laboratory
setup (section~\ref{sec:SimTurb}). To demonstrate the power of this method we
simulate a 1-km long free-space OAM-based communication link; these results are
presented in section~\ref{sec:Exp}.

\section{Simulating thick turbulence in the laboratory}\label{sec:SimTurb}
To examine the effects that a thick horizontal turbulent path might have on
OAM-based communication channel while still allowing information transfer and
AO correction, we chose to consider a 1\,km path $L$ with aperture sizes of the
sender and receiver of $D=18.2\,\mathrm{cm}$, at a wavelength of
$\lambda=785\,\mathrm{nm}$ corresponding to a Fresnel number of $N_f=\pi
D^2/(4\lambda L)=33$. The numerical value of $N_f$ is related to the total
number of spatial modes that the channel supports, and the value of $33$ was
chosen to be both realistic as well as large enough to support at least a few
dozen OAM modes~\cite{Mirhosseini2013}.

To allow for such a system to be realized in a laboratory setting, one must
find a way to incorporate turbulence into the channel, as well as properly
scale the system down to more manageable length scales. Section~\ref{sec:Turb}
contains a heuristic description of atmospheric turbulence and its various
effects on beam propagation. Section~\ref{sec:2PhaseModel} describes a method
of emulating a thick turbulence path with 2 thin phase screens that can be
represented in the lab with spatial light modulators (SLMs).  Scaling rules
that are invariant under Fresnel propagation are detailed in
section~\ref{sec:FresnelScaling} and the experimental setup is discussed in
\ref{sec:ExperimentalSetup}.

    \subsection{Atmospheric turbulence}\label{sec:Turb}
    Turbulence is a phenomenon that occurs in any fluid that is characterized
    by a large Reynolds number, $\mathcal{R}=V\mathfrak{L}/\nu$, where $V$ is the fluid's
    mean velocity, $\mathfrak{L}$ is the length scale of the fluid, and $\nu$ is the
    viscosity. The fluid will then break up into a cascade of turbulent eddies
    of decreasing size until the length scale, $\mathfrak{L}=l_0$ is such that
    $\mathcal{R}\le 1$ and the kinetic energy can be dissipated as heat. $l_0$
    is known as the inner scale and is typically on the order of 1\,mm in the
    atmosphere, negligibly small relative to the length scales in a free-space
    communication channel~\cite{Quirrenbach2006}.

    The fluid velocity $v(r)$ at any point $r$ in a turbulent fluid is a random
    process whose spatial structure can be described by the structure function,
    defined as $D_v(\mathbf{r_1,r_2})\equiv
    \left<\left|v(\mathbf{r_1})-v(\mathbf{r_2})\right|^2\right>$. For
    $\|\mathbf{r_1-r_2}\|>l_0$ the structure function is governed by
    ``Kolmogorov statistics,'' given by
    \begin{equation}
        D_v = C_v^2\delta r^{2/3},
        \label{eqn:Dv}
    \end{equation}
    where $\delta r \equiv \|\mathbf{r_1-r_2}\|$, and where $C_v^2$ is known as
    the velocity structure parameter which characterizes the strength of
    the fluctuations~\cite{Kolmogorov1961}. The turbulent eddies will mix the
    air, creating pockets of slightly different temperatures and thus
    pressures. As a consequence of this variation in air pressure the index of
    refraction will also vary statistically in the same way, leading to the
    refractive index structure function~\cite{Tatarski1961}
    \begin{equation}
        D_n = C_n^2\delta r^{2/3}.
        \label{eqn:Dn}
    \end{equation}

    The random fluctuations in the index of refraction is characterized by
    $D_n$ and lead to a random phase, $\phi(\mathbf{r})$ at the receiver. It
    was shown by Fried~\cite{Fried1965} that for Kolmogorov turbulence the
    phase structure function can be given by
    \begin{equation}
        D_\phi = 6.88 \left(\frac{\delta r}{r_0}\right)^{5/3},
        \label{eqn:Dphi}
    \end{equation}
    where $r_0$ is again Fried's parameter and can be calculated from $C_n^2$
    by the formula
    \begin{equation}
        r_0 = \left[\frac{2.91}{6.88} k^2 \int_0^L C_n^2(z) \,\mathrm{d}z\right]^{-3/5}
            = \left[\frac{2.91}{6.88} k^2 C_n^2 L\right]^{-3/5},
        \label{eqn:r0}
    \end{equation}
    where we have assumed for simplicity a constant value for $C_n^2$ along the
    path.

    A random phase imprinted on an optical beam will, upon propagation, will
    lead to variations in the amplitude as well, thus modifying the field at
    the receiver in both the phase and the amplitude. We can write the field as
    $U(\mathbf{r}) = U_0(\mathbf{r})\exp{\left(i\phi(\mathbf{r})+\chi
    (\mathbf{r})\right)}$, where $U_0$ is the field in the absence of
    turbulence, $\phi$ is the random phase, and $\chi$ is the random
    log-amplitude. These amplitude fluctuations, or ``scintillations'' can then
    be characterized by the variance in $\chi$ which is calculated from $C_n^2$
    as~\cite{Tatarski1961}
    \begin{equation}
        \sigma_\chi^2 = 0.563 k^{7/6} \int_0^L C_n^2(z) z^{5/6}\,\mathrm{d}z
            = \frac{0.563*6}{11} k^{7/6} C_n^2 L^{11/6}.
        \label{eqn:scintillation}
    \end{equation}

    Not only does $\chi$ cause scintillation within the beam, but this also
    leads to fluctuations in the total power of the beam, even if there are no
    losses in the path itself. These power fluctuations are due to beam wander
    and clipping by the finite aperture at the receiver. The normalized power
    over the aperture $\Sigma$ of area $A_\Sigma$, defined by $P\equiv
    \frac{1}{A_\Sigma}
    \int_\Sigma\,\mathrm{d}\mathbf{r}\exp{(2\chi(\mathbf{r}))}$, is used to
    numerically find the normalized power variance in the aperture,
    \begin{equation}
        \sigma_P^2 \equiv \left<P^2\right> - \left<P\right>
            = \left<P^2\right> - 1
            = \frac{1}{A_\Sigma^2}\iint_{\Sigma \Sigma'}\,
            \left(e^{4C_\chi(\mathbf{r},\mathbf{r'})} - 1 \right)\,\mathrm{d}\mathbf{r} \,\mathrm{d}\mathbf{r'} - 1,
        \label{eqn:PowerVariance}
    \end{equation}
    where $C_\chi(\mathbf{r},\mathbf{r'}) =
    \left<\left(\chi(\mathbf{r})-\left<\chi\right>\right)\left(\chi(\mathbf{r'})-\left<\chi\right>\right)\right>$
    is the log-amplitude covariance function~\cite{Andrews2005}. 

    \subsection{Two phase screen model}\label{sec:2PhaseModel}
    We have found that through use of 2 phase screens we can accurately model
    the horizontal turbulent channel, faithfully reproducing all of it's
    relevant statistical properties. Specifically we require this path to have
    the same values for $r_0$, $\sigma_\chi^2$, and $\sigma_P^2$ as described
    in section~\ref{sec:Turb}.   

    We can represent these 3 parameters by using thin Kolmogorov phase screens
    each with its own value of $r_0$.  The values of $r_0$ for each screen, as
    well as each screen's position along the path, give us 4 independent
    parameters that can be tuned until the 2 screen path reproduces the same 3
    parameters of the full horizontal channel. This still gives us an
    additional degree of freedom, and since phase vortices in $\phi$ create
    significant problems for AO correction as well as for OAM based
    encoding~\cite{Primmerman1995,Fried1992,Fried1998,Tyler2000}, we choose the
    density of branch points, $\rho_{BP}$, as our fourth parameter to constrain
    the problem.

    \begin{figure}[htpb]
        \centering
        \includegraphics[height=0.3\textheight]{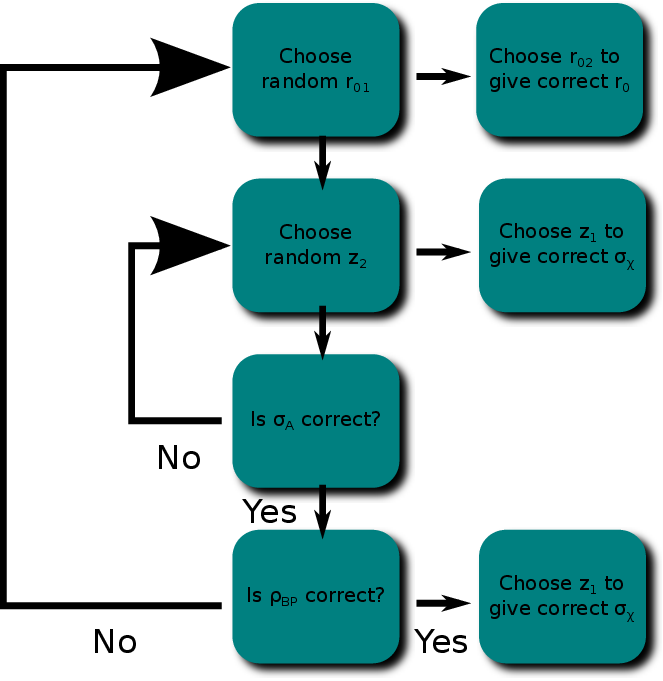}
        \caption{Procedure used to compute Fried's parameters and positions of
            the two thin phase screens needed to reproduce $r_0$,
            $\sigma_\chi$, $\sigma_p$, and $\rho_{bp}$ of the equivalent thick
            channel.}
        \label{fig:flowchart}
    \end{figure}

    For a horizontal path with a constant value of the refractive index
    structure parameter $C_n^2=1.8\times 10^{-14}\,\mathrm{m}^{-2/3}$, which
    represents a typical horizontal path near ground level, we compute the
    parameters given in Eqs.~\ref{eqn:r0}-\ref{eqn:PowerVariance}. The computed
    values are $r_0=24.4\,\mathrm{mm}$, $\sigma_\chi^2 = 0.197$, and
    $\sigma_P^2 = 7.04\times10^{-3}$. By Monte-Carlo simulation the density of
    branch points in $\phi$ is found to be $\rho_{BP} = 500/\mathrm{m}^2$.
    Physically this means that a typical realization of turbulence will create
    13 phase vortices within the receiver's aperture, which one could imagine
    being a serious impediment to one's ability to measure the intended phase
    vortex of the original transmitted OAM state.

    The second step in designing the two-phase-screen model is to find the
    values for the position and $r_0$ for each screen that will give the same
    values of $r_0$, $\sigma_\chi$, $\sigma_p$, and $\rho_{bp}$ of the thick
    path.  The procedure is diagrammed in Fig.~\ref{fig:flowchart}.  One starts
    with an initial guess for $r_{01}$ (i.e. $r_0$ for screen one), and then
    solves for the value of $r_{02}$ that will give the correct value for
    $D/r_0$.  Next, one randomly picks a value for the position of the second
    screen, $z_2$, and finds $z_1$ such that one gets the correct value for
    $\sigma_\chi$.  $z_2$ is then varied (along with $z_1$ to maintain
    $\sigma_\chi$) to set the correct value for $\sigma_P$.  Given this
    solution, $\rho_{BP}$ is computed by Monte-Carlo simulation.  If at this
    point we get the correct $\rho_{BP}$, a solution has been found; otherwise
    one starts over with a new choice for $r_{01}$.  Using this procedure we
    found we could simulate our 1\,km path with the parameters
    $r_{01}=3.926\,\mathrm{cm}$, $r_{02}=3.503\,\mathrm{cm}$,
    $z_1=171.7\,\mathrm{m}$ and $z_2=1.538\,\mathrm{m}$ (measured from the
    sender's aperture).

    \begin{figure}[htpb]
        \centering
        \includegraphics[width=0.7\textwidth]{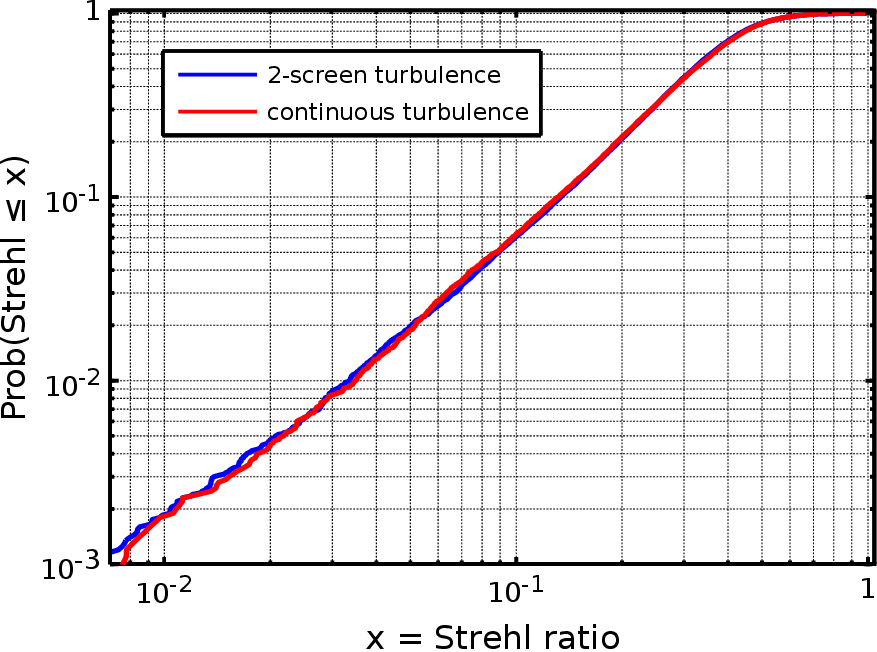}
        \caption{Cumulative distribution function (CDF) of the Strehl ratio for
            a thick turbulent path represented by 10 Kolmogorov phase screens
            (red line) and its equivalent 2 phase screen solution (blue line).}
        \label{fig:2vs12screenStrehl}
    \end{figure}


    As an independent test of this solution, a beam propagation simulation was
    performed to compare a thick turbulent path with the analogous 2-screen
    solution.  The continuous path was simulated using a standard split-step
    method in which the path $L$ is broken up into $N$ discrete steps. Each of
    the $N$ sections of turbulent atmosphere is replaced by non-turbulent
    propagation followed by an effective thin random phase screen that
    represents the effects of refractive index fluctuations within the slab.
    The propagation through a slab can be approximated by a thin screen so long
    as the scintillation due to propagation after encountering a random phase
    is negligible. As a rule of thumb one must require that the scintillation
    due to propagation through the slab must be less than 10\% of the total
    amount~\cite{Martin1988} in order to be able to represent the slab by a
    single screen, which is quantified as
    \begin{equation}
        \sigma^2_\chi (L/N) < 0.1 \sigma^2_\chi (L).
    \end{equation}
    We choose $N=10$, which for the horizontal path considered here becomes by
    Eq.~\ref{eqn:scintillation} 
    \begin{equation}
        \sigma^2_\chi (L/N)/\sigma^2_\chi = N^{-11/6} \approx 0.01 \ll 0.1.
    \end{equation}

    In each simulation a different random realization of turbulence was made
    and the Strehl ratio, defined as the ratio of the peak intensity to ideal
    peak intensity of a spot at a focal plane of the receiver, was computed.
    By repeating this many times, a probability distribution for the Strehl
    ratio was found and the results are shown in
    Fig.~\ref{fig:2vs12screenStrehl}.  As can be seen in the plot, the Strehl
    ratios of the 2 screen and the `continuous,' 10 screen paths show very good
    agreement with each other.  This result demonstrates that the 2 screen
    model not only reproduces the correct mean values for the statistical
    parameters of interest (by construction), but can also be expected to give
    similar distributions of possible measurement outcomes.

    \subsection{Fresnel scaling}\label{sec:FresnelScaling}
    The second thing that must be done to effectively simulate a turbulent path
    in a laboratory setting is to scale the optical paths down to more
    manageable lengths.  In order to ensure that the scaled path still
    represents the desired physical path, the propagation must remain invariant
    under the scaling.
    \begin{figure}[htpb]
      \centering
        \includegraphics[width=0.5\textwidth]{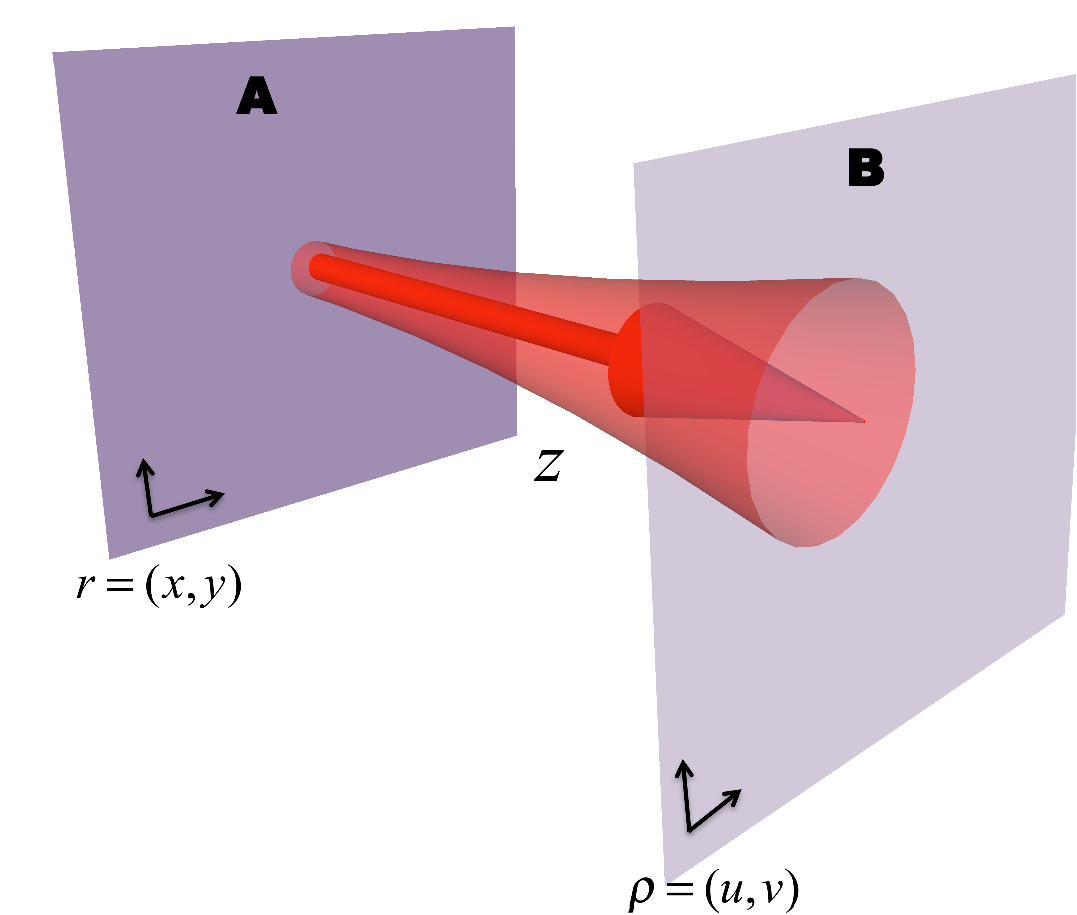}
        \caption{Propagation from plane A, described by coordinates $r=(x,y)$,
        to plane B with coordinates $\rho=(u,v)$}
        \label{fig:FresnelPropagation}
    \end{figure}
    Fresnel propagation from one plane to another a distance $z$ away as shown
    in Fig.~\ref{fig:FresnelPropagation} is given by:
    \begin{equation}
        U_B(\rho) = \frac{ie^{ikz}}{\lambda z}\int{U_A({r})e^{\frac{i\pi}{\lambda z}\left( {r} - \rho \right)^2}\, d^2{r}}
    \end{equation}
    Now if we scale the coordinates using ${r}'=\alpha_r{r}$,
    $\rho'=\alpha_\rho\rho$, and $z'=\alpha_z z$ then the above propagation
    equation becomes:
    \begin{equation}
        U_B\left(\frac{\rho'}{\alpha_\rho}\right) = 
        \frac{ie^{ikz'/\alpha_z}}{\alpha_r\lambda z'/\alpha_z}
        \int{U_A\left(\frac{{r}'}{\alpha_r}\right)e^{\frac{i\pi\alpha_z}{\lambda z'}\left( \frac{{r}'}{\alpha_r} - \frac{\rho'}{\alpha_\rho}\right)^2}\, d^2{r}'}
        \label{eqn:scaledProp}
    \end{equation}
    Since we require the Fresnel number to remain constant, $\alpha_z =
    \alpha_\rho\alpha_r$.  Then we can rewrite equation~\ref{eqn:scaledProp} as:
    \begin{equation}
        \frac{e^{ikz'\left( 1-1/\alpha_z \right)}}{\alpha_\rho}U_B\left(\frac{\rho'}{\alpha_\rho}\right)=
        e^{-i\frac{\pi\rho'^2}{\lambda f_\rho}}\left(
        \frac{ie^{ikz'}}{\lambda z'}
        \int{U_A\left(\frac{{r}'}{\alpha_r}\right) e^{-i\frac{\pi{r}'^2}{\lambda f_r}}e^{\frac{i\pi}{\lambda z'}\left( {r}' - \rho' \right)^2}\, d^2{r}'}
        \right)
        \label{eqn:compensatedScaledProp}
    \end{equation}
    where $f_\rho=z'/\left( 1-\frac{\alpha_r}{\alpha_\rho} \right)$ and
    $f_r=z'/\left( 1-\frac{\alpha_\rho}{\alpha_r} \right)$.  From
    Eq.~\ref{eqn:compensatedScaledProp} we see that the horizontal path between
    planes $A$ and $B$ can be scaled down (to within a scaling and phase
    constant) simply by adding a lens with focal length $f_r$ at $A$,
    propagating a distance $z'$, and then adding another lens with focal length
    $f_\rho$ at $B$ to cancel out the residual quadratic phase.

    \subsection{Experimental setup}\label{sec:ExperimentalSetup}
    \begin{figure}[htpb]
        \centering
        \includegraphics[width=0.9\textwidth]{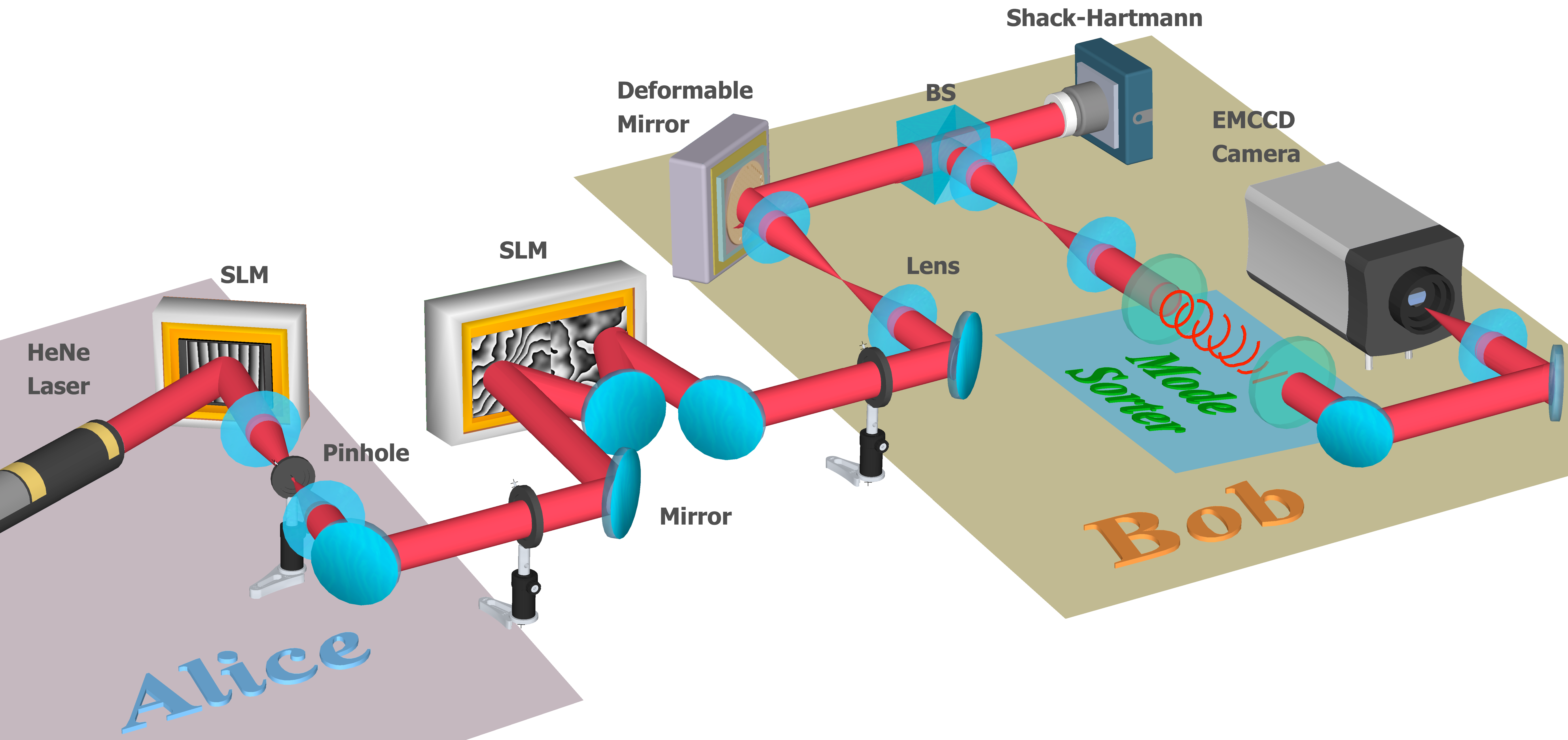}
        \caption{ Alice sends a beam prepared in a specific OAM state, $\ell$,
            to Bob.  Bob receives the beam after propagation through a channel
            representing a 1\,km turbulence path.  The beam is (optionally)
            corrected using a deformable mirror and sent to a sorter to make a
            measurement of the of the OAM spectrum of the beam.}
        \label{fig:ExperimentalSetup}
    \end{figure}
    A diagram of our experimental setup is presented in Fig.
    \ref{fig:ExperimentalSetup}.
    The sender, Alice, prepares a beam in a specific OAM state to send to the
    receiver, Bob, from an attenuated HeNe using a spatial light modulator
    (SLM) combined with a $4f$ system.  Diffraction gratings with spatially
    modulated properties can be shown to create any complex spatial field
    distribution in the first diffracted order (selected by the $4f$
    filter)\cite{Arrizon2007}. This applies whether the grating is an
    amplitude\cite{Mirhosseini2013a,Rodenburg2014a} or
    phase~\cite{Gruneisen2008} only hologram. A phase-only SLM was used in our
    experiment to allow for maximum efficiency in generating the modes.  The
    prepared state is then sent through the simulated 1\,km path scaled down as
    described in section \ref{sec:FresnelScaling} to a total length of
    1.3\,m.  The two thin phase screens used to simulate thick turbulence in
    our setup (section \ref{sec:2PhaseModel}) were implemented using an SLM in
    a double-pass configuration.  In addition, the quadratic phases required
    for proper scaling of the propagation path were added to the phases on the
    SLMs.
    
    After propagation through the turbulent channel, the beam at Bob's aperture
    is imaged with a $4f$ system onto a Thorlabs adaptive optics (AO) kit
    consisting of a $12\times 12$ actuator deformable mirror and a
    Shack-Hartmann wavefront sensor.  After the AO system, the beam is
    similarly imaged onto the first element (R1) of the OAM sorter.  The OAM
    sorter uses two refractive optical elements (R1 and R2) and a Fourier
    transforming lens (FT Lens) to spatially separate different OAM modes
    allowing for the OAM spectrum to be efficiently measured as described
    in~\cite{Lavery2012}.

\section{Experimental results}\label{sec:Exp}
In order to examine the effects of the turbulent channel on OAM communication,
we experimentally measured the OAM spectrum that Bob detects conditioned on
what Alice sent.  In a perfect channel, if Alice sends OAM mode $s$, then Bob
will measure an OAM spectrum that is simply a Kronecker delta centered at the
same mode.  However, in an imperfect or turbulent channel, there will be some
spreading into neighboring OAM modes to the prepared state.  The conditional
probability matrix, $P(d|s)$ where $d$ is the detected OAM mode and $s$ is the
sent mode, provides a natural expression for this crosstalk induced by the
imperfections or turbulence in the channel.

\begin{figure}[htpb]
    \centering
    \includegraphics[width=\textwidth]{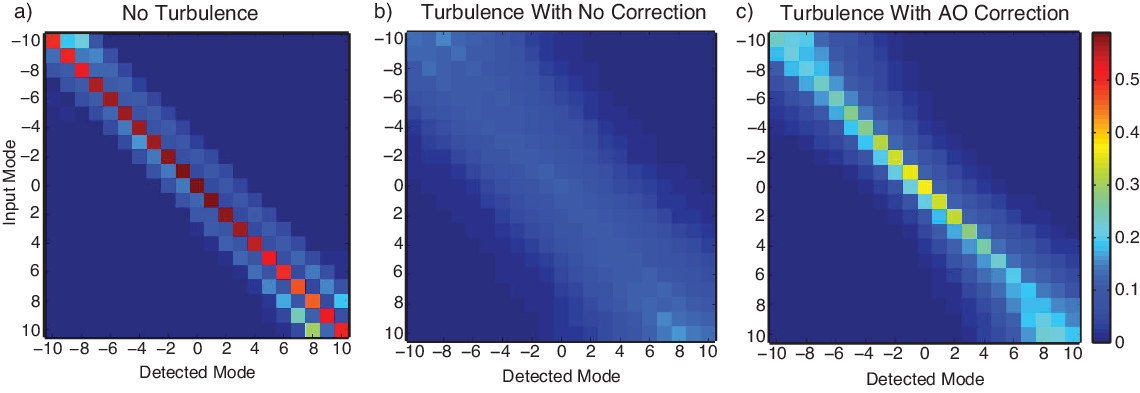}
    \caption{Measurement of the crosstalk in the channel represented by the
    conditional probability matrix, $P(d|s)$ for three cases: a) no turbulence,
    b) with turbulence, and c) with turbulence and adaptive correction.}
	\label{DataResults}
\end{figure}

$P(d|s)$ is plotted for 3 different scenarios in Fig.~\ref{DataResults}.
Fig.~\ref{DataResults}a shows $P(d|s)$ when there is no turbulence in the
channel, showing only crosstalk due to any misalignment in the system and
inherent crosstalk of the sorter~\cite{Lavery2012}.  Fig.~\ref{DataResults}b
shows the effects of thick turbulence ensemble averaged over 100 realizations,
which act to greatly spread the signal over many neighboring channels.  This
selective spreading of OAM into neighboring modes rather than randomly into any
OAM state is qualitatively similar to what is seen in the thin turbulence
regime demonstrated in~\cite{Rodenburg2012}.  For the third case shown in
Fig.~\ref{DataResults}c, adaptive correction was applied to the turbulence with
the AO system, allowing much of the signal to be recovered.  The phase
aberrations induced from each realization of turbulence was sensed and
corrected by the AO using the OAM $\ell=0$ mode.  Each mode was then sent
through the channel and AO system, and the OAM spectrum was measured by Bob.
This procedure was repeated and averaged over 50 realizations of turbulence.

\begin{figure}[htpb]
    \centering
    \includegraphics[width=7 cm]{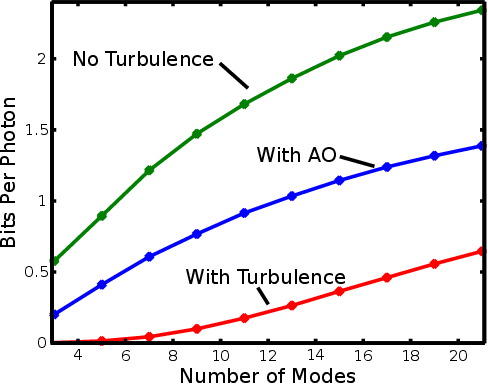}
    \caption{Measured channel capacity as a function of the number of OAM modes
        used for the spatial encoding.} 
    \label{fig:MIVsDim}
\end{figure}

In order to quantify the crosstalk induced by the turbulence as well as the
quality of the AO correction, we compute the mutual information between Alice
and Bob for all three cases above.  The mutual information between Alice and
Bob, $I(A;B)$, provides a measure of the channel capacity, as it gives the
maximum possible transmission rate that can be extracted by Bob in bits per
photon (or classically per detected pulse).  The mutual information is given by
the expression,
\begin{equation}
    I(A;B) = \frac{1}{N}\sum_{s,d}{P(d|s)\,\log_2{\left(\frac{P(d|s)N}{\sum_s P(d|s)}\right)}},
    \label{eqn:MI}
\end{equation}
where $N$ is the number of distinct symbols.  Fig.~\ref{DataResults}
qualitatively shows that thick turbulence greatly degrades the quality of the
channel.  Using Equation~\ref{eqn:MI}, we quantify these results by calculating
the mutual information as a function of the encoding dimension $N$.  The mutual
information for the three cases of no turbulence, thick turbulence, and
turbulence with AO correction, is plotted in Fig.~\ref{fig:MIVsDim} as a
function of $N$.  One can see that the AO system allows us to cancel roughly
half of the loss of channel capacity due to turbulence.

\begin{figure}[htpb]
    \centering
    \includegraphics[width=7 cm]{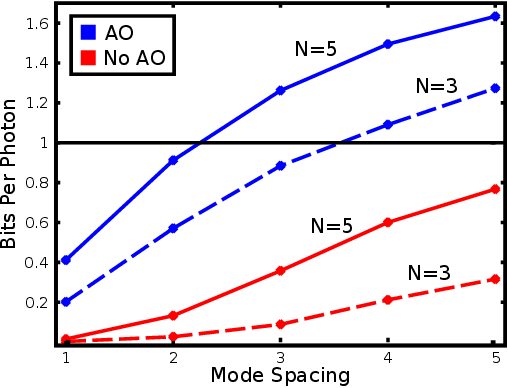}
    \caption{Measured channel capacity as a function of the spacing between OAM
        modes used for communication.}
	\label{fig:MIVsSpacing}
\end{figure}

Further, since turbulence preferentially scatters power into neighboring OAM
modes rather than randomly into all modes, one can increase the channel
capacity by choosing to use a less dense set of OAM modes~\cite{Malik2012}.
For instance, rather than encoding every OAM mode, one can encode in every
second mode (or third or fourth mode etc.).  Changing the encoding is also
independent of any AO system one may use, and thus a modified encoding can be
used along with AO correction to further enhance the channel capacity.
Fig.~\ref{fig:MIVsSpacing} shows the increase in the mutual information one can
obtain for a given number of encoded modes, $N$.  The channel capacity of an
ideal 2-bit system is shown for reference.  It is worth noting that the use of
spatial mode encoding shows an improvement over such a system with very
moderate resources (i.e. 3 modes with a channel spacing $>4$).

\section{Conclusions}
In this work we have demonstrated how one can experimentally simulate a thick
horizontal turbulent path as part of a free-space communication channel.  This
model was applied to the case of a 1\,km path and the effects were studied in
the context of an OAM communications channel.  It was shown that although the
turbulence severely degraded the channel, a high information capacity was still
possible by optimizing the encoding and adaptively correcting for some of the
induced phase aberrations.

We acknowledge Miles Padgett, Martin Lavery, and Daniel Gauthier for helpful
discussions.  Our work was supported by the Defense Advanced Research Projects
Agency (DARPA) InPho program and OSML also acknowledges support from the
CONACyT.

\section*{References}
\bibliographystyle{njp}

\bibliography{References}{}


\end{document}